\documentstyle[12pt,fleqn]{article}
\def\la{\langle}
\def\ra{\rangle}
\def\soc{{\rm C}_{60}}
\def\rug{{\rm C}_{70}}
\def\nex{N_{\rm ex}}

\def\beeq{\begin{equation}}
\def\eneq{\end{equation}}
\def\beeqa{\begin{eqnarray}}
\def\eneqa{\end{eqnarray}}

\addtocounter{section}{-1}
\begin{document}

\begin{center}
{\bf{ELECTRONIC STRUCTURES IN C$_{\bf 60}$-POLYMERS}} 

\mbox{}

{\rm Kikuo Harigaya\footnote[1]{E-mail address: 
harigaya@etl.go.jp; URL: http://www.etl.go.jp/People/harigaya/}}

\mbox{}

{\sl Physical Science Division, Electrotechnical Laboratory,\\ 
Umezono 1-1-4, Tsukuba, Ibaraki 305, Japan}

\end{center}

\mbox{}

\noindent
Variations in the band structures of $\soc$-polymers are 
studied, when conjugation conditions and the electron number 
are changed.  We use a semiempirical model with the 
Su-Schrieffer-Heeger type electron-phonon interactions.  
In the neutral one-dimensional $\soc$-polymer, electronic 
structures change among direct-gap insulators and the metal, 
depending on the degree of conjugations.  High pressure 
experiments could observe such pressure-induced metal-insulator 
transitions.  The $\soc$-polymer doped with one electron per 
one molecule is always a metal.  The energy difference 
between the highest-occupied state and the lowest-unoccupied 
state of the neutral system becomes smaller upon doping owing 
to the polaron effects.  When the $\soc$-polymer is doped 
with two electrons per one $\soc$, the system is insulating.  
When the conjugation in the direction of the polymer 
chain is smaller, it is a direct-gap insulator.  The energy 
gap becomes indirect when the conjugation is stronger.
We also study the antiferromagnetic phase of $A$$\soc$ 
by using a tight-binding model with long-range Coulomb 
interactions.  The antiferromagnetism is well described 
by the model.  The comparison with the photoemission 
studies shows that the new band around the Fermi energy 
of the $A$$\soc$ phase can be explaind by the extremely 
large intrusion of an energy level into the gap of 
the neutral system.  This indicates that the interaction 
effects among electrons are important in doped $\soc$-polymers.

\mbox{}

\begin{center}
{\bf INTRODUCTION}
\end{center}

\mbox{}

Recently, it has been found that the linear (one-dimensional) 
$\soc$-polymers are realized in alkali-metal doped $\soc$ crystals: 
$A\soc$ ($A=$K, Rb, Cs) [1-5], and their solid state properties 
are intensively investigated.  One electron per one $\soc$ is 
doped in the polymer chain.  It seems that Fermi surfaces exist 
in high temperatures, but the system shows antiferromagnetic 
correlations in low temperatures [1].  The $\soc$-polymer has 
lattice structures where $\soc$ molecules are arrayed in a linear 
chain.  The bonds between $\soc$ are formed by the [2+2] cycloaddition 
mechanism.  The structures of the one-dimsensional polymers are 
shown in Fig. 1.  There are four membered rings between neighboring 
$\soc$ molecules.

In this article, we shall review studies on variations in the 
band structures of the one-dimensional $\soc$-polymers when the 
conjugation conditions between molecules are changed.  We assume 
that the bonding states between $\soc$ might be changed easily 
possibly by applying high pressures.  As discussed in ref. [2], 
there are three candidates for the classical bonding 
structures around the four membered ring.  The first case (A) 
is that there are only weak van der Waals interactions between 
$\soc$, and the classical double bonds remain as in the isolated 
$\soc$ molecule.  The second case (B) is that all the bonds of 
the four membered ring are the single bond, so they are 
$\sigma$-like in the quantum chemistry.  All the four-membered 
rings in Fig. 1 are formed by the $\sigma$-bonds in this 
case.  Then, the third case (C) is that the bonds which connect 
the neighboring molecules are the double bonds, and the bonds 
derived from the isolated $\soc$ -- the bonds, $\langle 1,2 \rangle$ 
and $\langle 3,4 \rangle$, of Fig. 1 -- are destroyed completely.  
Here, $\langle i,j \rangle$ 
indicates the pair of the neighboring $i$ and $j$th atoms.
In this case, the bonds between $\soc$ -- the bonds, 
$\langle 1,3 \rangle$ and $\langle 2,4 \rangle$, of Fig. 1 -- have 
the $\sigma$- as well as $\pi$-characters.  In other words, the 
degree of the conjugations between the molecules becomes 
maximum.  We have proposed a model which can deal with 
changes of conjugation conditions among the above three cases, 
and have looked at electronic band structures of  
$\soc$-polymers [6,7].  We note that the operator at the lattice 
sites of the four membered rings is one of the relevant linear 
combinations of $sp^3$ orbitals, assuming a possibility of local 
$\sigma$-conjugations at the four membered rings.  The similar 
assumption of the $\sigma$-conjugation has been used in Si-based 
polymers, for example, in ref. [8].  We, however, use the term 
``conjugation" for simplicity in this article, because the local 
$\sigma$-conjugations can be regarded as a part of the global 
conjugations which are extended over the system.

In refs. [6,7], we have proposed a semiempirical tight-binding 
model analogous to the Su-Schrieffer-Heeger (SSH) model [9] of 
conjugated polymers.  The first purpose of this article is to
review the model and the main results.  The electronic structures 
may depend sensitively upon the conjugation conditions even in 
the neutral polymer, because the several bonds connecting neighboring 
molecules are largely distorted, and the mixings among $sp^3$ 
orbitals will change only by slight change of the bond structures [2].  
We have studied effects of the change of the conjugation conditions 
by introducing a phenomenological parameter in a tight-binding 
model.  The model is an extension of the SSH-type model which
has been applied to $\soc$ [10,11] and $\rug$ [11,12] molecules. 
The model is solved with the assumption of the adiabatic approximation,
and band structures are reported in order to discuss metal-insulator 
changes and polaron effects by varying conjugations.

We have concluded that the electronic structures change among 
direct-gap (in\-di\-rect-gap) insulators and the metal, depending on 
the degree of conjugations.  In the neutral polymer, the electronic 
structures change from the direct gap insulator with the gap 
at the $\Gamma$ point, through the metal, to the insulator 
with the direct gap at the Brilloune zone boundary, as increasing 
the conjugation between $\soc$ molecules.  The high pressure 
experiments may be able to change conjugation conditions 
between $\soc$ molecules, and the electronic structure changes 
could be observed.  The reentrant change from the insulator 
through a metal into an insulator is specific to the present 
$\soc$-polymer systems and is quite interesting.

We will also review the doping effects in one-dimensional polymers [13].
We discuss the following properties.  (1)  The $\soc$-polymer 
doped with one electron per one molecule is always a metal.  
The energy difference between the highest-occupied state and 
the lowest-unoccupied state of the neutral system becomes smaller 
upon doping owing to the polaron effects.  (2) When the 
$\soc$-polymer is doped with two electrons per one $\soc$, 
the system is insulating.  When the conjugation in the direction 
of the polymer chain is smaller, it is a direct-gap insulator.  
The energy gap becomes indirect when the conjugation is stronger.

The last purpose of this article is to review a microscopic model 
for the antiferromagnetism in one-dimensional $\soc$-polymers [14] 
in order to show that the antiferromagnetic ground state can be actually 
explained by our model with strong electron-electron interactions.  
We have pointed out the importance of Coulomb interactions, comparing 
with photoemission experiments of $\soc$-polymer systems [15].
The model is one dimensional and has two contributions.  
The first term is the hopping interactions along the polymer backbone 
structure, and the second term is the Coulomb interaction among 
electrons.  We use the long-range Ohno potential which we have used 
in the previous paper [16].  In ref. [17], the relevance of the 
Hubbard model, where the $\soc$ molecule is regarded as a site, 
has been discussed.  We will compare the present calculation with 
the assumption of the Hubbard model.

\mbox{}

\begin{center}
{\bf MODELS}
\end{center}

\mbox{}

We first apply an SSH-type model to $\soc$-polymers.  In $\soc$, 
all the carbon atoms are equivalent, 
so it is a good approximation to neglect the mixing between $\pi$- 
and $\sigma$-orbitals.  The presence of the bond alternation and 
the energy level structures of the neutral $\soc$ molecule can be 
quantitatively described by the calculations within the adiabatic 
approximation.  In $\rug$, the molecular structure becomes longer, 
meaning that the degrees of the mixing between $\pi$- and 
$\sigma$-characters are different depending on carbon sites.  
In this respect, the extended SSH model does not take account of 
the differences of the mixings.  However, it has been found [11,12] 
that qualitative characters of the electronic level structures are 
reasonably calculated when the extended SSH model is applied to 
the $\rug$.  This is a valid approach because the $\sigma$-orbitals 
can be simulated by the classical harmonic springs in the first 
approximation.

In this article, we assume the same idea that the lattice 
structures and the related molecular orbitals of 
each $\soc$ molecule in the $\soc$-polymers can be described 
by the SSH-type model with the hopping interactions for the 
$\pi$-orbitals and the classical springs for the $\sigma$-orbitals.  
However, the mixings between the $\pi$- and $\sigma$-orbitals
near the eight bonds, $\langle i,j \rangle$ ($i,j=1 - 4$),
shown in Fig. 1 are largely different from those of regions
far from the four bonds.  We shall shed light on this special
character of bondings between the neighboring $\soc$.
Electronic structures would be largely affected by changes
of conjugation conditions (or local $\sigma$-conjugations
as in Si-based polymers [8]) around the four bonds.
We shall introduce a semiempirical parameter $a$ as shown
in the following hamiltonian:
\beeqa
H_{\rm pol} &=&  a \sum_{l,\sigma} 
{\sum_{\langle i,j \rangle}}^{'} 
(- t + \alpha y_{l,\langle i,j \rangle} )
( c_{l,i,\sigma}^\dagger c_{l+1,j,\sigma} + {\rm h.c.} )  \nonumber \\
&+&  (1-a) \sum_{l,\sigma} 
{\sum_{\langle i,j \rangle}}^{''} 
(- t + \alpha y_{l,\langle i,j \rangle} )
( c_{l,i,\sigma}^\dagger c_{l,j,\sigma} + {\rm h.c.} )  \nonumber \\
&+& \sum_{l,\sigma} \sum_{\langle i,j \rangle = {\rm others}}
(- t + \alpha y_{l,\langle i,j \rangle} )
( c_{l,i,\sigma}^\dagger c_{l,j,\sigma} + {\rm h.c.} ) \nonumber \\
&+& \frac{K}{2} \sum_i \sum_{\langle i,j \rangle} y_{l,\langle i,j \rangle}^2,
\eneqa
where $t$ is the hopping integral of the system without the 
bond alternations in the isolated $\soc$ molecule; $\alpha$ is the 
electron-phonon coupling constant which changes the hopping 
integral linearly with respect to the bond variable 
$y_{l,\langle i,j \rangle}$, where $l$ means the $l$th
molecule and $\langle i,j \rangle$ indicates the pair of
the neighboring $i$ and $j$th atoms; the atoms with $i=1-4$
of the one-dimensional polymer are shown in Fig. 1;
the other $i$ and $j$ in the third column of eq. (1) label 
the nonnumbered atoms in the same molecule; $c_{l,i,\sigma}$ 
is an annihilation operator of the $\pi$-electron 
at the $i$th site of the $l$th molecule with spin $\sigma$; 
the sum is taken over the pairs of neighboring atoms;
and the last term with the spring constant $K$ 
is the harmonic energy of the classical spring simulating the 
$\sigma$-bond effects.  Note that the sum with the prime is 
performed over $\langle i,j \rangle = \la 1,3 \ra$ and $\la 2,4 \ra$.
The sum with the double prime is performed over 
$\la i,j \ra = \la 1,2 \ra$ and $\la 3,4 \ra$.

As stated before, the parameter $a$ controls the strength of 
conjugations between neighboring molecules.  When $a=1$,
the $\sigma$-bondings between atoms, 1 and 2, 3 and 4, 
in Fig. 1 are completely broken and the orbitals  would 
become like $\pi$-orbitals.  The bond between the atoms 
1 and 3 and the equivalent other bonds become double bonds.  
This is the case (C).  As $a$ becomes smaller, the conjugation 
between the neighboring molecules decreases, and the $\soc$ 
molecules become mutually independent.  In other words, the 
interactions between molecules become smaller in the intermediate $a$ 
region.  In this case (B), the operator $c_{l,i,\sigma}$ at the lattice 
sites of the four membered rings is one of the relevant linear 
combinations of the $sp^3$ orbitals.  Here, we assume a 
possibility of local $\sigma$-conjugations at the four membered rings.
When $a=0$ [the case (A)], the $\soc$ molecules are completely 
isolated each other.  The band structures of the $\soc$-polymers 
will change largely depending on the conjugation conditions.  
The effects are reviewed in the next section.

We further discuss the mechanism of the antiferromagnetism
in one-dimensional $\soc$-polymers doped with one electron per
one $\soc$.  The model is the following:
\beeqa
H &=& H_{\rm 0} + H_{\rm int}, \\
H_{\rm 0} &=&  - at \sum_{l,\sigma} 
\sum_{\langle i,j \rangle = \langle 1,3 \rangle,\langle 2,4 \rangle} 
( c_{l,i,\sigma}^\dagger c_{l+1,j,\sigma} + {\rm h.c.} ) \nonumber \\
&-&  (1-a)t \sum_{l,\sigma} 
\sum_{\langle i,j \rangle = \langle 1,2 \rangle,\langle 3,4 \rangle} 
( c_{l,i,\sigma}^\dagger c_{l,j,\sigma} + {\rm h.c.} )  \nonumber \\
&-& t \sum_{l,\sigma} \sum_{\langle i,j \rangle = {\rm others}}
( c_{l,i,\sigma}^\dagger c_{l,j,\sigma} + {\rm h.c.} ), \\
H_{\rm int} &=& U \sum_{l,i} 
(c_{l,i,\uparrow}^\dagger c_{l,i,\uparrow} - \frac{n_{\rm el}}{2})
(c_{l,i,\downarrow}^\dagger c_{l,i,\downarrow} 
- \frac{n_{\rm el}}{2}) \nonumber \\
&+& \sum_{l,l',i,j} W(r_{l,l',i,j}) 
(\sum_\sigma c_{l,i,\sigma}^\dagger c_{l,i,\sigma} - n_{\rm el})
(\sum_\tau c_{l',j,\tau}^\dagger c_{l',j,\tau} - n_{\rm el}).
\eneqa
In eq. (2), the first term is the tight binding part of the
$\soc$-polymer, and the second term is the Coulomb interaction
potential among electrons.  The eq. (3) is equivalent to eq. (1),
if $\alpha = 0$.  The electron-phonon interactions are not 
taken into account, because they are not necessary in order to show 
the presence of the antiferromagnetism.  The numbers in eq. (3) 
indicate the carbon atoms displayed in Fig. 1.  Equation (4) is the Coulomb 
interactions among electrons.  Here, $n_{\rm el}$ is the number 
of electrons per carbon site; $r_{l,l',i,j}$ is the distance 
between the $i$th site of the $l$th $\soc$ and $j$th site of 
the $l'$th $\soc$; and
\beeq
W(r) = \frac{1}{\sqrt{(1/U)^2 + (r/r_0 V)^2}}
\eneq
is the Ohno potential.  The quantity $W(0) = U$ is the strength 
of the onsite interaction; $V$ means the strength of the long 
range part; and $r_0 = 1.433$\AA~ is the mean bond length of the 
single $\soc$ molecule.

\mbox{}

\begin{center}
{\bf DOPING EFFECTS AND ELECTRONIC STATES\\
IN ONE-DIMENSIONAL C$_{\bf 60}$-POLYMERS}
\end{center}

\mbox{}

We use the parameters, $t=2.1$eV, $\alpha = 6.0$eV/\AA, and 
$K = 52.5$eV/\AA$^2$, which give the energy gap 1.904eV and 
the difference between the short and bond lengths 0.04557\AA\ 
for an isolated $\soc$ molecule.

Figure 2 shows the excess electron distribution for the three
conjugation conditions, $a=0.5$, 0.8, and 1.0.  
The labels of sites, A-I, are shown in Fig. 1.  Due to the 
reduced symmetry of the polymer chain, mutually symmetry equivalent 
sites have the same electron density.  The each label represents 
the site with the different electron density.  The white bars 
are for the case of the excess electron number per $\soc$, 
$\nex = 1$, and the black bars are for $\nex=2$.  
The excess electron density at the sites A is the largest for all 
the displayed cases.  The bond alternation patterns are largely 
distorted near these sites, so the electron density change is 
the largest too.  In Fig. (a), the densities at sites, D, F, 
and H, are relatively larger.  In Figs. (b) and (c), the densities 
are larger at sites D and I.  In this way, the sites, where 
excess electrons prone to accumulate and thus the bond alternation
patterns are highly distorted, are spatially localized in the 
molecular surface.  This is one of the polaron effects, which 
we have discussed in ref. [11].  Here, we do not show 
bond alternation patterns for simplicity.  We only note that the 
distortion of the bond alternation is larger where the change of 
the electron density is larger.  The polaronic distortion pattern 
is different from that in the isolated $\soc$, and this is owing 
to the difference in the symmetry group.  Tanaka et al [17] have
drawn a schematic figure where the electron density change
is the largest at the molecule center, but the present result 
does not agree with this feature.  Numerical calculations
are necessary in order to derive the actual distributions.

Next, we discuss band structures of electrons in detail.
Figures 3-5 display the band structures for the conjugation
parameters, $a=0.5$, 0.8, and 1.0, respectively.  Figures (a),
(b), and (c) are for $\nex=0$, 1, and 2, respectively.  In each
figure, the unit cell is taken as unity, so the first Brilloune
zone extends from $-\pi$ to $\pi$.  Due to the inversion symmetry,
only the wavenumber region, $0 \leq k \leq \pi$, is shown in the figures.

Figures 3 (a-c) show the band structures of the polymer for
$a=0.5$ and with $\nex=0$, 1, and 2, respectively.  In Fig. 3(a),
the highest fully occupied band is named as ``HOMO", and the lowest 
empty band as ``LUMO".  There is an energy gap about 0.8 eV at the 
zone center.  The system is a direct gap insulator.  When doped 
with one electron per $\soc$, the system is a metal as shown
by the presence of the Fermi surface in Fig. 3(b).  The system
is an insulator again when $\nex=2$, as shown in Fig. 3(c).
Here, the energy gap is at the boundary of the Brilloune zone,
i.e., at $k = \pi$.

As increasing the parameter $a$, the overlap of the HOMO band 
and LUMO band appears in the neutral system.  This is shown 
for $a=0.8$ in Fig. 4(a).  There are Fermi surfaces, so the 
system changes into a metal.  If $a$ increases further, the 
positions of the previous HOMO band and LUMO band are reversed 
as shown for $a = 1.0$ in Fig. 5(a).  The system becomes a 
direct gap insulator again.  The energy gap is at $k=\pi$.

When $\nex = 1$, the system is always a metal when $a$ varies.
The representative cases, $a = 0.8$ and 1.0, are displayed in
the Figs. 4(b) and 5(b).  The number of the Fermi surface is
two or four, depending upon the parameter $a$.  However, the 
metallic property is obtained for all the $a$ we take.
We also find that the HOMO band and LUMO band of the neutral
system shift into the energy gap upon doping.  The positions
of the other energy bands do not change so largely.
This is due to the polaronic distortion of the lattice,
which we have discussed in the calculation of an isolated
molecule in ref. [11].

When $\nex = 2$, the system turned out to be always an
insulator.  For smaller $a$, for example, $a = 0.5$ and 0.8,
the energy gap appears at $k=\pi$.  For larger $a$, for example,
$a = 1.0$ [Fig. 5(c)], the energy gap becomes an indirect gap.
The polaronic distortion becomes larger as the doping 
concentration increases.  Thus, the intrusions of the HOMO 
and LUMO bands of the neutral system become larger, too.

The above variations of the energy gap are summarized 
for the cases $\nex = 0$ and 2, where a finite energy 
gap appears for a certain $a$ value.  The results are 
shown in Figs. 6(a) and (b).  The white (black) squares 
indicate that the system is a direct gap insulator where 
there is a energy gap at $k=0$ ($\pi$).  The squares with 
the plus mean that the system is an indirect gap insulator.  
The crosses are for metals.  In the neutral system 
$\nex = 0$ [Fig. 6(a)], the energy gap decreases almost 
linearly for smaller $a$.  The system changes into a metal 
as $a$ increases, and finally an energy gap appears again.   
For $\nex = 2$ shown in Fig. 6(b), the system is a direct gap 
insulator with the energy gap at $k = \pi$ up to $a \sim 0.9$.  
The system turns into an indirect gap insulator near $a = 1.0$.

\mbox{}

\begin{center}
{\bf MAGNETISM IN ONE-DIMENSIONAL C$_{\bf 60}$-POLYMERS}
\end{center}

\mbox{}

The moldel eq. (2) is solved with the assumptions of the 
unrestricted Hartree-Fock approximation.  We assume that 
one electron is doped per one $\soc$ molecule and $V=U/2$.  
The results are shown by changing the parameters, $a$ and $U$.

The magnitude of the ordered spin per one $\soc$ is calculated,
and is shown in Fig. 7 as a function of $a$ and $U$.  If the magnitude 
is zero, there is not an antiferromagnetic order.  The magnitude
is finite when the magnetic order appears.  As $U$ becomes stronger, 
the magnetization appears and becomes larger.  This is a natural 
consequence.  When $a$ is smaller, the itinerancy of electrons 
in the chain direction decreases, and thus the antiferromagnetism 
appears more easily.  This feature is realized in the present 
calculation: the critical value of $U$, where a finite magnetization 
begins to appear, decreases as $a$ becomes smaller.

The calculated results are summarized as a phase diagram in Fig. 8. 
The metallic phase region is named as M.  And the antiferromagnetic
phase is named as AF.  The antiferromagnetism appears in the small 
$a$ region when $U$ is taken constant.  The phase boundary between two 
phases is an almost linear line.  In ref. [17], the applicability of the
one-dimensional Hubbard model is discussed.  When this discussion
is adopted to the present model eq. (2), the model can be mapped
to the following hamiltonian:
\beeq
H = - t_{\rm eff} \sum_{\la l,l' \ra,\sigma}
(a^\dagger_{l,\sigma} a_{l',\sigma} + {\rm h.c.})
+ U_{\rm eff} \sum_{l} n_{l,\uparrow} n_{l,\downarrow},
\eneq
where $t_{\rm eff}$ is an effective hopping integral between 
neighboring $\soc$ molecules; $U_{\rm eff}$ is the on-ball
Coulomb repulsion; $a_{l,\sigma}$ is an annihilation operator
of an effective orbital of the doped electron at the $l$th molecule;
and $n_{l,\sigma} = a_{l,\sigma}^\dagger a_{l,\sigma}$.  
In the one-dimensional Hubbard model, it is known that the
antiferromagnetism appears even when $U_{\rm eff}$ is small but
positive.   But, there is a parameter region where a magnetic 
order is not present in Fig. 8.  Thus, we find that the
mapping onto the Hubbard model is not always relevant to
all the parameter set.  The mapping could be used for the
case with small $a$ and large $U$.  In ref. [17], the 
antiferromagnetism has been discussed in connection with 
the polaron in the doped $\soc$ [11].  In the present
section, electron-phonon interactions are not considered, and only
electron-electron interactions are taken into account. 
Therefore, the antiferromagnetism is not directly related with the 
polaron formation due to the Jahn-Teller effects [11].  Rather,
we could regard the antiferromagnetism as one of the spin 
density wave states in the itinerant electron systems.
It is of course that the doped extra electrons tend to
have large amplitudes at certain sites on the surface of
the $\soc$ molecules.  However, the main origin of the 
localization would be the strong electron-electron 
interactions rather than the Jahn-Teller effects.

It is useful to look at how the band structures change
upon the formation of the antiferromagnetic order.  The 
dispersions of the bands of the system before the electron 
doping and without the magnetic order are shown for 
$(a,U,V) = (0.3, 2.4t, 1.2t)$ in Fig. 9.  Here, the electron 
number is 60 per one $\soc$, and the unit cell consists 
of one $\soc$.  The band structures of the antiferromagnetic
system are shown for the same $(a,U,V)$ in Fig. 10.
The unit cell becomes doubled, so the first Brilloune zone
is half of that of Fig. 9.  The energy bands in Fig. 10 become
narrower.  The band named ``HOMO" in Fig. 9 changes into the 
HOMO-1 band and the HOMO-2 band in Fig. 10.  There is a small 
energy gap between the HOMO-1 band and HOMO-2 band at the 
boundary of the first Brilloune zone in Fig. 10.  In the same 
way, the band named ``LUMO" of Fig. 9 is the origins of the 
``HOMO" and ``LUMO" of Fig. 10.  There is a large energy gap 
at the zone boundary as well.  These energy gaps are opened 
by the formation of the antiferromagnetism.

It should be noted that the energy band intrudes into the original 
energy gap of the neutral system extremely largely.  The HOMO band of 
the system with the antiferromagnetism even locates near the center 
of the original energy gap.  This intrusion is much larger than that
in the polaron formation due to the Jahn-Teller interactions
in the doped $\soc$ [11], and cannot be obtained from models 
with electron-phonon interactions only.  Actually, the new band
has been observed in the photoemission studies of the orthorhombic 
phase of Rb$\soc$ [15].  Such the large changes in band 
structures might be the evidence of strong electron-electron 
interactions in the $\soc$-polymers rather than the results of 
the Jahn-Teller distortion due to the electron-phonon interactions.
The positions of the HOMO band and HOMO-1 band of Fig. 9 do not 
shift so much upon the formation of the antiferromagnetism.
This is also consistent with the photoemission studies [15].

In the present calculations, we have not considered possible interchain 
interactions which might effect on the dispersion of bands
perpendicular to the polymer direction [18,19].  In the tight-binding 
type models as used in this article, there might be many 
candidates for the interchain interaction forms.  Even though they have 
not been taken into account in the present calculations, we 
might expect that the nature of one-dimensinal chains with 
strong electron-electron interactions is still effective on 
the antiferromagnetism in the real $A$$\soc$ systems 
because of the low dimensionality of the polymer-chain which
might be favorable for the orderings to occur due to interactions among 
itinerant electrons.

\mbox{}

\begin{center}
{\bf SUMMARY}
\end{center}

\mbox{}

In the neutral one-dimensional  
$\soc$-polymers, electronic structures change among direct-gap 
insulators and the metal, depending on the degree of 
conjugations.  The high pressure experiments may be able 
to change conjugation conditions in the chain direction, 
and the electronic structure changes could be observed.

The one-dimensional $\soc$-polymer doped with one electron 
per one molecule is always a metal.  The energy difference 
between the highest-occupied state and the lowest-unoccupied state 
of the neutral system becomes smaller upon doping owing to 
the polaron effects.  When the one-dimensional $\soc$-polymer 
is doped with two electrons per one $\soc$, the system is 
insulating.  When the conjugation in the direction of the polymer 
chain is smaller, it is a direct-gap insulator.  The energy 
gap becomes indirect when the conjugation is stronger.

We have studied the antiferromagnetic phase of the $\soc$-polymers
by microscopic calculations.  The antiferromagnetism is well described 
by the present model.  The formation of the magnetic order is closely
related with the strong interactions rather than the Jahn-Teller 
mechanisms.  We have compared with the photoemission studies to 
find that the new band around the Fermi energy of the $A$$\soc$ 
phase can be explaind by the fact that a energy level intrudes 
extremely largely into the gap of the neutral system by the electron 
doping.  This is another indication that the interaction effects 
among electrons play a crucial role in doped $\soc$-polymers.

\mbox{}

\begin{center}
{\bf REFERENCES}
\end{center}

\mbox{}

\noindent
$[1]$ O. Chauvet, G. Oszl\`{a}nyi, L. Forr\'{o}, P. W. Stephens,
M. Tegze, G. Faigel, and A. J\`{a}nossy, Phys. Rev. Lett. {\bf 72}, 
2721 (1994).\\
$[2]$ P. W. Stephens, G. Bortel, G. Faigel, M. Tegze,
A. J\`{a}nossy, S. Pekker, G. Oszlanyi, and L. Forr\'{o},
Nature {\bf 370}, 636 (1994).\\
$[3]$ S. Pekker, L. Forr\'{o}, L. Mihaly, and A. J\`{a}nossy,
Solid State Commun. {\bf 90}, 349 (1994).\\
$[4]$ S. Pekker, A. J\`{a}nossy, L. Mihaly, O. Chauvet,
M. Carrard, and L. Forr\'{o}, Science {\bf 265}, 1077 (1994).\\
$[5]$ M. Kosaka, K. Tanigaki, T. Tanaka, T. Atake, A. Lappas,
and K. Prassides, Phys. Rev. B {\bf 51}, 12018 (1995).\\
$[6]$ K. Harigaya, Phys. Rev. B {\bf 52}, 7968 (1995).\\
$[7]$ K. Harigaya, Chem. Phys. Lett. {\bf 242}, 585 (1995).\\
$[8]$ T. Hasegawa, Y. Iwasa, H. Sunamura, T. Koda, Y. Tokura,
H. Tachibana, M. Matsumoto, and S. Abe, Phys. Rev. Lett.
{\bf 69}, 668 (1992).\\
$[9]$ W. P. Su, J. R. Schrieffer, and A. J. Heeger, Phys. Rev. B
{\bf 22}, 2099 (1980).\\
$[10]$ K. Harigaya, J. Phys. Soc. Jpn. {\bf 60}, 4001 (1991).\\
$[11]$ K. Harigaya, Phys. Rev. B {\bf 45}, 13676 (1992).\\
$[12]$ K. Harigaya, Chem. Phys. Lett. {\bf 189}, 79 (1992).\\
$[13]$ K. Harigaya, Chem. Phys. Lett. (1996) (to be published).\\
$[14]$ K. Harigaya, Phys. Rev. B {\bf 53}, R4197 (1996).\\
$[15]$ G. P. Lopinski, M. G. Mitch, J. R. Fox, and J. S. Lannin,
Phys. Rev. B {\bf 50}, 16098 (1994).\\
$[16]$ K. Harigaya and S. Abe, Phys. Rev. B {\bf 49}, 16746 (1994).\\
$[17]$ K. Tanaka, Y. Matsuura, Y. Oshima, T. Yamabe,
Y. Asai, and M. Tokumoto, Solid State Commun. {\bf 93}, 163 (1995).\\
$[18]$ S. C. Erwin, G. V. Krishna, and E. J. Mele, Phys. Rev.
B {\bf 51}, 7345 (1995).\\
$[19]$ K. Tanaka, Y. Matsuura, Y. Oshima, T. Yamabe, H. Kobayashi, 
and Y. Asai, Chem. Phys. Lett. {\bf 241}, 149 (1995).\\

\pagebreak

\begin{flushleft}
{\bf Figure captions}
\end{flushleft}

\noindent
Fig. 1.  The crystal structure of the one-dimensional $\soc$-polymer.
The labels, A-I, indicate carbon atoms whose charge densities
are not equivalent due to the symmetry.  The conjugations along 
four bonds, which connect carbon atoms with labels, 1-4, are controled
by the parameter $a$ in eq. (1).

~

\noindent
Fig. 2.  Excess electron distribution for the three 
conjugation conditions, (a) $a=0.5$, (b) 0.8, and (c) 1.0.  
The white bars are for the case $\nex = 1$ and the 
black bars are for $\nex=2$.

\mbox{}

\noindent
Fig. 3.  Band structures of the $\soc$-polymer of the case $a = 0.5$.
The excess electron number per one $\soc$ is (a) 0, (b) 1, 
and (c) 2, respectively.  In (a), the highest fully occupied band is 
named as ``HOMO", and the lowest empty band as ``LUMO".
The lattice constant of the unit cell is taken as unity.

\mbox{}

\noindent
Fig. 4.  Band structures of the $\soc$-polymer of the case $a = 0.8$.
The excess electron number per one $\soc$ is (a) 0, (b) 1, 
and (c) 2, respectively.  The lattice constant of the unit cell 
is taken as unity.

\mbox{}

\noindent
Fig. 5.  Band structures of the $\soc$-polymer of the case $a = 1.0$.
The excess electron number per one $\soc$ is (a) 0, (b) 1, 
and (c) 2, respectively.  In (a), the highest fully occupied band is 
named as ``HOMO", and the lowest empty band as ``LUMO".
The lattice constant of the unit cell is taken as unity.

\mbox{}

\noindent
Fig. 6.  The variations of the energy gap plotted against $a$.
The cases $\nex = 0$ and 2 are shown in (a) and (b), respectively.
The white (black) squares indicate that the system is
a direct gap insulator where there is a energy gap at
$k=0$ ($\pi$).  The squares with the plus symbol mean that the system
is an indirect gap insulator.  The crosses are for metallic cases.

\mbox{}

\noindent
Fig. 7.  The magnitude of the magnetization per one $\soc$
shown by changing $a$ and $U$.  We use $V = U/2$ here. 
The closed and open squares are for $a= 0.1$ and $0.2$,
respectively.  The closed and open circles are for
$a=0.3$ and $0.4$.

\mbox{}

\noindent
Fig. 8.  The phase diagram shown against $a$ and $U$.  
The region with ``M" is the metallic phase, and
the region with ``AF" is the antiferromagnetic phase.

\mbox{}

\noindent
Fig. 9.  The band structures of the neutral $\soc$-polymer 
of the case $(a,U,V)=(0.3, 2.4t, 1.2t)$.  The unit cell 
consists of one $\soc$.   The length of the unit cell is taken
as unity.  The highest fully occupied band is named as ``HOMO", 
and the lowest empty band as ``LUMO".

\mbox{}

\noindent
Fig. 10.  The band structures of the $\soc$-polymer doped with
one electron per one $\soc$ of the case $(a,U,V)=(0.3, 2.4t, 1.2t)$.  
The unit cell consists of two $\soc$ molecules.   The length of 
the unit cell is taken as unity.  The highest fully occupied band 
is named as ``HOMO", and the lowest empty band as ``LUMO".

\end{document}